\documentclass[a4paper,11pt]{article}
\pdfoutput=1
\usepackage{jheppub}

\usepackage{graphicx,caption,subcaption}
\usepackage{mathrsfs}
\usepackage{float}

\title{A holographic superconductor forced through interactions}

\author[a]{Pallab Basu}
\affiliation[a]{University of the Witwatersrand,1 Jan Smuts Ave, Johannesburg, 2000, South Africa}

\author[b]{, Jyotirmoy Bhattacharya}
\affiliation[b]{Department of Physics, Indian Institute of Technology
Kharagpur, Kharagpur 721302, India}

\author[b]{and Sayan Kumar Das}

\emailAdd{pallab.basu@wits.ac.za, jyoti@phy.iitkgp.ac.in, sayankumardas@iitkgp.ac.in}

\abstract{
We propose a novel mechanism to achieve superconductivity at zero chemical potential, within the holographic framework. 
Extending previous construction of the holographic superconductors, we consider an Einstein-Maxwell system coupled with two interacting scalars in Anti-de Sitter space.
One of the scalar fields is charged and therefore, interacts non-trivially with the gauge field, while the other is uncharged. 
We find that, if we turn on a boundary source for the uncharged scalar field, it forces the condensation of the charged scalar, leading to a superconducting phase in the dual 
boundary theory. The condensation occurs at a certain critical value of the source, depending on the value of the chemical potential, which can even be zero. 
We work out the complete phase diagram of this scenario. We further corroborate the existence of superconductivity at zero chemical potential, through a fluctuation analysis on our solution. 
Notably, the conductivity of the system, as a function of probing frequency, exhibits characteristics of usual holographic superconductors. We also investigate how 
these properties of the system changes, as we vary the interaction strength between the scalar fields. Our results indicate a controlled mechanism 
to manipulate the phase transition temperature of superconductors with strongly coupled microscopics.}

\keywords{gauge-gravity duality, holographic superconductors, strongly correlated systems, high $T_c$ superconductivity.}

\begin{document}

\begin{flushright} 
\small{ARXIV PREPRINT} 
\end{flushright}

\maketitle
\flushbottom
%

\section{Introduction and summary}\label{sec:intro}

One of the outstanding recent theoretical developments is the framework of gauge-gravity duality \cite{Maldacena:1997re}, which relates a 
theory of gravity to a gauge QFT in one lower dimension. This holographic principle provides us with a theoretical laboratory to study 
complex phenomenon in strongly coupled QFTs geometrically, with the help of a classical gravity dual. This duality has been exploited to 
model and analyse a large class of low-energy emergent phenomenon in strongly coupled systems, which are otherwise difficult to study directly. This idea has been used to address questions in condensed matter physics using holography \cite{Hartnoll:2016apf}.

In recent times, several unconventional superconductors has been discovered (such as the cuprates and other organic compounds), 
in which the mechanism leading to superconductivity is unclear. The most remarkable feature that characterises these unconventional 
superconductors is their relatively high transition temperatures ($T_c$). In these materials, 
it is clear that the conventional BCS theory fails to provide an adequate justification for the pairing mechanism involved in process 
of condensation, leading to the superconducting phase. An understanding of the true mechanism behind the increased $T_c$ is 
extremely important, since it may lead to the realisation of superconductors at room temperature through controlled manipulation.

Now, these systems which exhibit superconductivity at high $T_c$, have an underlying microscopics which is strongly coupled and there are 
limited theoretical tools to analyse them directly. However, recently developed holographic techniques, have been used to model 
systems with strongly coupled microscropics (see \cite{Hartnoll:2016apf}, for a review). 
The gravity system, through the holographic dictionary, describes a field theory with strongly coupled dynamics. This, therefore,
 gives us an excellent theoretical laboratory to test and experiment with new phenomenon which occur due to strong coupling and understand the mechanisms behind them. Our lessons 
from this exercise, may provide us with new insights in to the functioning of real life high $T_c$ superconductors, such as 
the cuprates. 
Although the details of the microscopic dynamics may be different in the cuprates compared to the system described by gravity, it is expected that they share some universal 
features, which owes its existence to strongly coupled dynamics. Here we seek a qualitative understanding of such universal features. Our set up would not be very useful for any quantitative 
prediction about the cuprates.
%

Following a lead by \cite{Gubser:2008px}, the holographic duals of superconductors and superfluids 
were constructed in \cite{Hartnoll:2008kx,Hartnoll:2008vx}. This development was followed by early generalizations \cite{Franco:2009yz,Chen:2010hi,Pan:2010at} 
and an extensive work and in this area by many authors (see \cite{Hartnoll:2016apf,Herzog:2009xv,Horowitz:2010gk,Sachdev:2010ch} and 
the references therein). 

In \cite{Hartnoll:2008kx}, an Einstein-Maxwell system was considered, in the presence of charged massive scalar. 
This system admits two dominant finite temperature phases relevant for 
our discussion here. The first phase is that of an RN blackhole, with the scalar field being zero, which we shall refer to as the normal phase. 
In the second phase, the charged scalar is non-zero and the solution is that of a charged black hole with a scalar hair. 
The non-zero regular bulk scalar field configuration, leads to a non-zero vev of the dual scalar operator 
in the boundary theory. Thus from the point of view of boundary theory,  this corresponds to 
a phase where a bosonic charged scalar operator has condensed (and hence has a non-zero vev) which is characteristic of a superconductor
{\footnote{ Strictly speaking, the $U(1)$ symmetry is global in nature and the corresponding gauge fields are not 
dynamical in the boundary theory. For this reason, this construction describes a superfluid \cite{Basu:2008st,Herzog:2008he}. 
A similar set up has also been successfully explored to study superfluids in the hydrodynamic approximation, 
and their holographic duals are slowly  fluctuating hairy black holes (for instance, see \cite{Bhattacharya:2011eea,Bhattacharya:2011tra}). 
However, most of the phenomenon characteristic of superconductors are also observed in this system, as we may consider 
the $U(1)$ symmetry to be weakly gauged. For instance, it has been demonstrated in \cite{Hartnoll:2008kx} that in this system, 
currents obeying London equations are generated, which is a hallmark of superconductivity. Hence, this system 
can be reliably used to study phenomenon related to superconductivity. 
%
}}.

In all the discussions of the holographic superconductor so far, the presence of a non-zero chemical potential was extremely important. 
The phase transition from the normal to the superconducting phase, was essentially driven by the interaction of the charge carriers with the gauge field. 
In the holographic set up, the non-zero boundary chemical potential, resulting in a non-trivial profile of the gauge field in the bulk, provided the necessary 
effective potential, for the charged scalar to condense into a symmetry broken phase. In other words, the boundary chemical potential, above a critical value, was the key ingredient 
that ensured the existence of a dynamical instability of the normal phase, the end point of which was superconductivity. Holographic principle maps bulk $U(1)$ gauge symmetry to a $U(1)$ global symmetry in the boundary and the corresponding charge may be thought of as a density. In this particular respect, the mechanisim of holographic supercondcutor is similar to the bose condensation in 
a  free bosonic theory, 
where also density is extremely important in effecting the phase transition. 

In this note, we report a novel mechanism 
to achieve superconductivity, {\it {even at zero chemical potential}}. Here, we study an extension of the set up of \cite{Hartnoll:2008kx} to include two interacting scalar fields. 
One of our scalar fields corresponds to 
the charged scalar field in \cite{Hartnoll:2008kx}, but for us, the other scalar field is uncharged. However, the nontrivial interaction between the 
charged and uncharged scalar fields leads to a very interesting and important consequence with regards to the phase structure of the normal and superconducting phases. 
It has been already observed earlier \cite{Basu:2010fa, Musso:2013rnr, Bigazzi:2011ak}, that interacting bulk scalars fields leads to significant and novel phenomenon in 
holographic superconductors. Our generalisation here is strongly motivated by \cite{Chaturvedi:2014dga} (also see \cite{Musso:2013rnr}), where in similar holographic system of two interacting scalars (without any $U(1)$ symmetry), it was possible to make one of the scalars to condense, just by turning on the source for the other scalar. 

Thus following \cite{Chaturvedi:2014dga}, in our set-up, we turn on a boundary source $J$ for the uncharged scalar field. 
We find that, irrespective of the value of the chemical potential $\mu$ including zero, 
the condensation of the charged scalar takes place at sufficiently high value of the source of the uncharged scalar field. 
In other words, through a particular kind of forcing (which, in this case, is the 
source for the uncharged scalar operator), we are able to facilitate the process of condensation of the charged scalar field, thus creating a superconducting phase. 

In the absence of any such forcing $J=0$, our results reduce to the results of \cite{Hartnoll:2008kx}, where the superconducting phase starts existing at a finite 
critical value of the chemical potential $\mu_c$, at a fixed temperature. Now as we introduce forcing with the source $J$, this critical value of the chemical 
potential reduces, with the increase in $J$. In fact, in this way, a critical value of $J$ is reached when $\mu_c$ tends to zero. 
This is our main result and is summarised in the phase diagram in fig.\ref{fig:phpl}.

The holographic system that we study has an underlying conformal symmetry. Due to this, although we perform our analysis at a fixed temperature, 
solutions with other values of temperature are related to our solutions through a conformal transformation. So, in our set-up, instead of 
considering the length scales associated with temperature and chemical potential separately, we should consider the 
dimensionless ratio between the two. In particular, the relevant critical parameter is $\mathcal T_c = T_c / \mu_c$. 
This critical parameter $\mathcal T_c$, increases with the increase of the forcing $J$. 
Thus, if we were to change the temperature of the system keeping the chemical potential fixed, we would have found 
that the transition temperature $T_c$, increases as we increase $J$. Hence, forcing the system with $J$, provides us with a controlled mechanism of increasing $T_c$. 
Therefore, it would be extremely interesting to identify such additional interacting operator, in a real life superconductors. 

Besides presenting the solutions and the phase diagram, we have also performed several related analysis of this system. 
We are able to confirm that the nature of the phase transition is sensitive to the form of interactions between the two scalar fields. The phase transition 
is generically second order. But, when the uncharged scalar is linearly coupled with the charged scalar 
and when their mutual interactions are stronger than their self-interactions, 
we can even see a first order phase transition in this system. 
We come to this conclusion, by carefully analysing the free-energies of the competing phases
\footnote{ Note that, in our case, 
the two competing phases are both hairy back holes (in the probe approximation). The normal phase in our case corresponds 
to an uncharged scalar hair of RN black hole, with the charged scalar field being zero. 
This is because, while comparing free-energies, we must compare solutions with the same value of boundary sources.}.

We also perform a linear fluctuation analysis of our solutions, to compute the conductivity of our system. 
We confirm that, we indeed have a  superconducting phase, particularly 
in the limit when the chemical potential tends to zero.  The conductivity of the system as function of probing frequency, also exhibits a soft gap 
\footnote{Here, by gap we mean that, there is a critical value of the frequency $\omega_c$, below which the conductivity is very close to zero (with a delta function at zero frequency). 
The vanishing of conductivity below $\omega_c$, indicates that there is a gap in the spectrum of excitations (see \cite{Horowitz:2010gk} for more details). 
Now, for any generic thermal system, as long as we are at non-zero temperature, the conductivity, although small, never completely vanishes below $\omega_c$.
For the usual holographic superconductor, below $\omega_c$,  the conductivity does not completely vanish, even in the zero temperature limit \cite{Horowitz:2009ij}. 
For this reason, the gap is referred to as a soft gap for holographic superconductor studied in \cite{Horowitz:2009ij}. 
In this paper, we have not explored the zero temperature limit, but we expect the gap to remain soft in our case as well, since 
our set up is very similar to that analysed in \cite{Horowitz:2009ij}. In order to obtain a hard gap, a more complex holographic set up is necessary, for instance see \cite{Basu:2009vv}.}, 
characteristic of holographic superconductors.


\section{ Holography of  forced condensation of a charged scalar operator}\label{sec:mainbody}

As explained in \S \ref{sec:intro}, one of our main objective here, is to realize the superconducting phase, holographically, at zero chemical potential. 
We therefore, consider a gravitational system, which can reliably  
describe a superconductor with strongly coupled microscopics
\footnote{As explained in \S \ref{sec:intro}, 
although, the details of the field theory 
in such a holographic description, differ from those describing the microscopics of real life high $T_c$ superconductors, they are expected to capture most of the 
essential macroscopic qualitative features. 
%
%
}. 
For simplicity and concreteness, we shall consider superconductivity in $2+1$ dimensions. 
The mechanism and the phenomenon we describe here, are expected to be robust, if we increase the number of space-time dimensions. 
The holographic system would therefore be a $3+1$ dimensional system, which we now proceed to describe. 

\subsection{The holographic system}\label{ssec:setup}

We work with the Einstein-Maxwell system with two interacting scalar fields. The action for our system is given by
\begin{equation}\label{pac}
\begin{split}
\mathcal S_{\text{EM}} = \int d^4x ~ & \sqrt{-g} \bigg( R + \Lambda + \left( -\frac{1}{4} F_{\mu \nu}F^{\mu \nu} - (\mathcal D_\mu \psi )(\mathcal D_\mu \psi )^* - m_1^2 |\psi|^2 
- \frac{\alpha_1}{2} |\psi|^4   \right. \\
&  \left. - (\partial_\mu \phi )(\partial^\mu \phi ) - m_2^2\phi^2 - \frac{\alpha_2}{2} \phi^4 + \lambda \phi |\psi|^2 + \beta \phi^2 |\psi|^2
\right)  \bigg)
\end{split}
\end{equation}
where $\mathcal D_\mu =\nabla_\mu - i A_\mu$, is the gauge covariant derivative, while $\nabla_\mu$ is simply the covariant derivative in curved space. Since we wish 
to work within the framework of standard $AdS/CFT$ correspondence, we shall consider a negative cosmological constant $\Lambda$. As is clear from the action, 
the complex scalar field $\psi$ is charged under the gauged $U(1)$ symmetry, with $A_\mu$ being the gauge field and $F_{\mu \nu}$ being the corresponding 
field strength. The other scalar field $\phi$ is uncharged. 

Most of the popular constructions of the holographic superconductors, used a set-up similar to \cite{Hartnoll:2008kx}, but without any uncharged scalar field like $\phi$. In the 
absence of $\phi$, a non-zero chemical potential, above a certain critical value, is absolutely crucial to sustain a superconducting phase. This is because, 
in the presence of the chemical potential, we have a non-trivial profile of the gauge field in the bulk. This, in turn, modifies the effective potential 
for the charged scalar field $\psi$, in such a way that  $\psi$ acquires a negative effective mass (below the BF bound in AdS). This results in an instability 
of the normal phase, for all chemical potentials above the critical value. The end point of this instability is superconductivity. So, in such a scenario, 
the mechanism that leads to superconductivity, is solely driven by the concentration of the charge carriers. 

In our work here, we are able to achieve the superconducting phase, even at zero chemical potential. 
 The main novelty in our construction, is the uncharged scalar $\phi$, which interacts with the charged scalar $\psi$. 
If we now turn on a source $J$ for the uncharged scalar $\phi$, the mutual interactions 
between the scalars, is now able to generate an effective potential for $\psi$, such that the normal phase becomes unstable. This also leads to
 the condensation of $\psi$, even when the bulk gauge field is zero, and hence we have a vanishing boundary chemical potential. 

Thus, we are proposing a completely new mechanism, which can lead to superconducting instabilities in a strongly 
coupled system. We shall also demonstrate that the superconducting phase exists, when the chemical potential, 
as well as the source $J$, are both non-zero, suggesting that the two mechanisms can operate simultaneously, and in fact, 
they complement one another.

\subsubsection*{Details of the set-up}\label{ssec:setup}

In this note, we shall work in the probe approximation, in which the contribution of the matter part of the Lagrangian is considered small 
\footnote{This approximation is justified, when the charge of the scalar field is large, and we work with suitable normalizations for the matter fields. Also it has been noted before that finite temperature physics of holographic superconductor does not change significantly with gravity backreaction. \cite{Hartnoll:2008vx}}.
Under this approximation, the leading order equations of motion for the metric are the vacuum Einstein equations, 
with the negative cosmological constant.  A finite temperature solution of the vacuum Einstein equations, is the AdS-Schwarzschild black hole whose metric is given by 
\begin{equation*} \label{bhbg}
\begin{split}
& ds^2=\frac{1}{u^2}\left(-f(u)dt^2+\frac{1}{f(u)}du^2+dx^2+dy^2 \right),~~\text{where}~~f(u)=(1-u^3)
\end{split}
\end{equation*}
Here, $u=1$ is the horizon radius, which is proportional to the temperature of the black hole. As explained in \S \ref{sec:intro}, our system has 
an underlying conformal symmetry, which makes all temperature length scales equivalent. Thus, without loss of any generality, we can exploit 
this feature and set $u =1$ as the horizon. This is a useful simplification, particularly for the numerical calculations and so we shall use this 
throughout our analysis. 


By considering the metric to be of the form \eqref{bhbg}, the Einstein equations are solved once and for all. In the background \eqref{bhbg},
we shall consider the dynamics of the gauge field and the interacting scalars, without considering any back reaction on the metric. 
This approximation, is not expected to affect the generality of our results in any crucial way. 

For the gauge field, we shall choose to work in the radial gauge $A_r = 0$. This choice is compatible with the ansatz of considering the 
charged scalar field to be real $\psi^* = \psi$.  In this gauge, using the reality condition 
on the charged scalar, the probe action of interest reduces to 
\begin{equation}\label{realpac}
\begin{split}
\mathcal S = \int d^4x ~ & \left( -\frac{1}{4} F_{\mu \nu}F^{\mu \nu} -   (\mathcal \partial_\mu \psi )(\mathcal \partial^\mu \psi ) - m_1^2 \psi^2 
- \frac{\alpha_1}{2} \psi^4   \right. \\
 & \left. - (\partial_\mu \phi )(\partial^\mu \phi ) - m_2^2 \phi^2 - \frac{\alpha_2}{2} \phi^4 - A^2 \psi^2 + \lambda \phi \psi^2 + \beta\psi^2 \phi^2
\right) 
\end{split}
\end{equation}
Notice that, besides considering mass term for the two scalar fields, we have also considered a quartic self interaction. The 
self interaction terms are somewhat important for supporting the solutions that we derive. There are two terms representing the mutual interaction between the two scalars, the 
strengths are given by $\beta$ and $\lambda$. 

For convenience of manipulation and a clear presentation of the results, we shall treat the two cases separately, one in which 
$\lambda =0$ and other in which $\beta =0$. No additional new features are expected to appear, 
if we turn on both $\beta$ and $\lambda$ simultaneously. Also, to facilitate numerical manipulations (see \S \ref{ssec:numsol} for more details), we shall fix the masses of both 
the scalar fields to be
\begin{equation} \label{mch}
m_1^2  = -2 = m_2^2.
\end{equation}
Note that although we have chosen 
a negative masses for both the scalar fields, it is higher than the the BF bound in $AdS_4$.

When, $\lambda = 0$, there is an upper bound on the value of $\beta$, arising from the requirement that the potential has to be bounded from below. This condition, 
is given by $\beta \leq \sqrt{\alpha_1 \alpha_2}$. There is also a lower bound on $\beta$ \cite{Chaturvedi:2014dga}, which arises from the existence criterion of the 
required solutions. 
%
%
This bound may be estimated as follows. When the charged scalar field is zero, the value acquired by the uncharged scalar field deep in the IR region of 
AdS, should be well approximated by the minima of its potential, which is given by $\phi^2 = - m_2^2 / \alpha_2$. The effective mass for $\psi$ in the IR, is therefore, 
$M_\psi = m_1^2 - \beta \phi^2 = m_1^2 - \beta  m_2^2 / \alpha_2$. As long as this mass is below the BF bound of AdS in 3+1 dimensions, we shall have the superconducting 
instability and therefore the solutions that we seek, is likely to exist. If we choose \eqref{mch}, this converts to the following bound on $\beta$
\begin{equation}
\beta \geq \frac{\alpha_2}{8}.
\end{equation} 

As we will see later in \S\ref{ssec:ordphtr}, for the set of allowed values of $\beta$, the nature of the phase transition is always second order. 
In \S\ref{app:toy}, using a toy model, we shall demonstrate that, the upper bound on $\beta$, is primarily responsible for not having a first order phase transition, 
when the mutual interaction is of this nature.

However, when $\beta = 0$, there is no upper bound for the interaction strength $\lambda$. This is because, the potential always remains bounded from below, 
as long as the quartic self-interaction of the scalar fields are non-zero. Therefore, in this scenario, it is possible to increase the value of $\lambda$, to values, comparable 
to the strength of self interactions. It is in this case, that we find that the order of the phase transition becomes first order (see \S\ref{ssec:ordphtr}). We provide an understanding of this 
through a toy model in \S\ref{app:toy}. 

Let us note, however, $\lambda$ has a lower bound, analogous to $\beta$, arising from the criterion of existence of superconducting instability 
in the system. Again, if we choose the masses to be \eqref{mch}, this bound may be estimated exactly as it was done for $\beta$ and we must have 
\begin{equation}
\lambda \geq \frac{\sqrt{\alpha_2}}{{4 \sqrt{2}}}.
\end{equation} 

The equation of motion that follows from the action \eqref{realpac} are 
\begin{equation} \label{reom}
\begin{split}
& \nabla_\mu F^{\mu \nu} - 2 A^\nu \psi^2 = 0 , \\
& \nabla_\mu \nabla^\mu \psi - A^2 \psi - m_1^2 \psi   - \alpha_1 \psi^3 + \lambda \phi \psi +\beta \phi^2 \psi = 0 ,\\
& \nabla_\mu \nabla^\mu \phi - m_2^2 \phi - \alpha_2 \phi^3 + \frac{\lambda}{2} \psi^2 + \beta \phi \psi^2 = 0 . 
\end{split}
\end{equation}

We shall now go on to discuss the solutions to these set of equations. 

\subsection{Numerical results in the probe approximation}\label{ssec:numsol}

We look for finite temperature static solutions, where the fields depend only on the radial coordinate of AdS. Also for most of 
our analysis we shall only turn on the time component of the gauge field, which would give us solutions with non-zero electric field 
in the bulk and no magnetic field. Thus we make the following ansatz for the probe matter fields 
\begin{equation} \label{eqan}
A = A_t(u) dt , ~\phi = \phi(u) , ~\psi=\psi(u). 
\end{equation}
With this ansatz the equation of motion \eqref{reom} reduces to 
\begin{equation}\label{expleq}
\begin{split}
& A_t''(u) u^2 f(u)+ 2 A_t (u) \psi^2(u)= 0 \\
& \psi''(u) u^2 f(u)+ \left( u^2 f'(u)-2 u f(u) \right) \psi'(r) -\alpha_1 \psi^3(u)+ \left(\beta \phi^2(u)+ \lambda \phi +\frac{A_t^2(u)}{f(u)}u^2+2 \right)\psi(u) = 0 \\
& \phi''(u) u^2 f(u)+ \left( u^2 f'(u)-2 u f(u) \right) \phi'(r) -\alpha_2 \phi^3(u)+ \left(\beta \psi^2(u)+2 \right)\phi(u) + \frac{\lambda}{2} \psi(u)^2= 0 \\
\end{split}
\end{equation}
%
%
These set of coupled equations are hard to solve analytically. Therefore, we have to resort to numerical computations. Before we proceed to describe 
the numerical computation, let us discuss the asymptotic behaviour of the fields involved and their boundary conditions. 

\subsubsection*{Boundary Conditions}

For our numerical analysis, it is particularly convenient to set the masses of the scalar fields to $m_1^2 = - 2 = m_2^2$.  This is because, with this choice of the masses 
the asymptotic behaviour of the fields near the boundary of $AdS$ is given by 
\begin{equation} 
\begin{split}
& \phi (r) =  J u + \langle \mathcal O _\phi \rangle u^2 +  \dots  \\
& \psi (r) =  J_\psi u + \langle \mathcal O _\psi \rangle u^2 +  \dots  \\
& A_t (r) = \mu - \rho u + \dots 
\end{split}
\end{equation}
Note that due to the specific choice of masses, the scalar fields do not diverge near the boundary and scales with integer powers of the radial 
coordinate \footnote{ This feature is particularly convenient while extracting the sources and operator expectation values
from the asymptotic behaviour of the scalar fields. This is the only reason, for choosing the special value of the masses for the scalar fields.}.

Here the coefficients of the asymptotic behaviour of the fields each have a direct interpretation from the boundary field theory point of view. 
 $ \langle \mathcal O _\phi \rangle$ is the vev of the boundary scalar operator dual to the uncharged scalar field $\phi$, 
 and  $J$ is the corresponding source of this dual operator. Similarly, $J_\psi$ and $\langle \mathcal O_ \psi \rangle$ are the source 
 and vev of the operator dual to the charged scalar field $\psi$. For the gauge field, $\mu$ corresponds to the boundary chemical potential, 
 while $\rho$ is interpreted as the boundary charge density.

\begin{figure}[H]
\centering
\begin{subfigure}{.5\textwidth}
 \centering
 \includegraphics[width=0.9\textwidth]{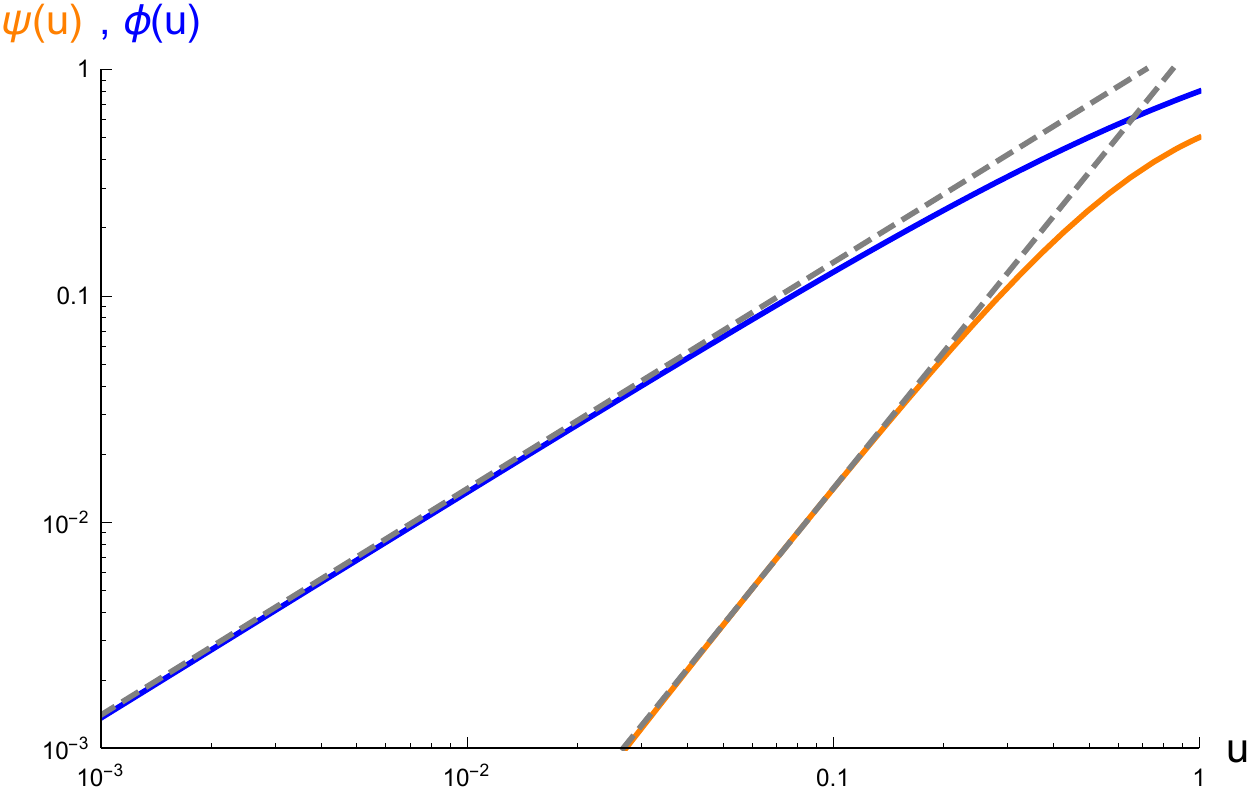}
 \caption{}
 \end{subfigure}%
\begin{subfigure}{.5\textwidth}
 \centering
 \includegraphics[width= 0.9\textwidth]{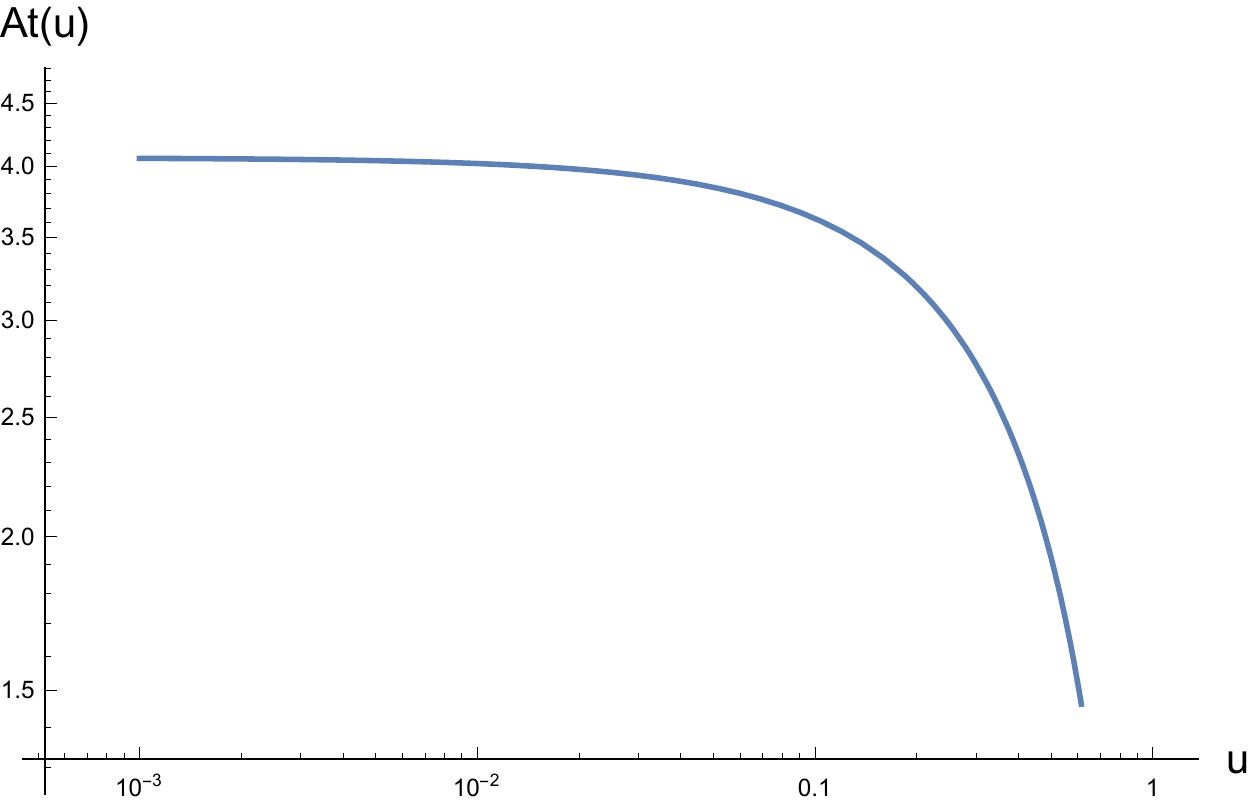}
 \caption{}
\end{subfigure}
\caption{A sample solution for $\alpha_1=1=\alpha_2$, $\beta=0.6$ and $\lambda =0$. The dashed lines in (a) are $u$ and $u^2$ respectively, appearing as straight lines in the Log-Log plot. 
Shooting has been performed to ensure that $\psi$ falls off as $u^2$,  
near the boundary $u\rightarrow 0$, thus ensuring $J_\psi = 0$. The gauge field goes to a constant near the boundary, with the value of the constant denoting the chemical potential.}\label{fig:samsol}
\end{figure}

Near the horizon, we require the scalar fields to be regular and the time component of the gauge field to vanish. The demand of the regularity 
of the scalar fields near the horizon may sound like an inconsistent assumption at first, particularly since the choice of coordinates for the metric is such 
that it blows up at the horizon. There is a coordinate singularity of the black hole in Schwarzschild coordinates. However, our requirement of the regularity of the scalar fields is justified by 
the fact that we would be working with a static ansatz. Since our fields do not depend on the time coordinate, the equations that we solve in the Schwarzschild coordinates
are going to be identical to those in future Eddinton-Finklestein coordinates, which has a regular future event horizon. This justifies our regular boundary conditions,
near the even horizon. This requirement is implemented by the following near horizon expansion for the fields 
\begin{equation} 
\begin{split}
& \phi (u) = \phi_0 + (1-u) \phi_1 + (1-u)^2 \phi_2 + \dots  \\
& \psi (u) =  \psi_0 + (1-u) \psi_1 + (1-u)^2 \psi_2 + \dots \\
& A_t (r) = -a_0 (1-u) + a_1 (1-u)^2 + a_2 (1-u)^3  \dots 
\end{split}
\end{equation}
If we plug this expansion, back into the equations \eqref{expleq}, the higher order coefficients $\phi_1, \phi_2, \psi_1, \psi_2, a_1 $ and $a_2$ are all determined in terms 
of the leading order coefficients $\phi_0$, $\psi_0$ and $a_0$. This gives us a series solution about the horizon, and is utilized in supplying the necessary IR boundary conditions slightly 
away from the horizon, for generating the numerical solutions. 

\begin{figure}[H]
\centering
\begin{subfigure}{0.5\textwidth}
 \centering
 \includegraphics[width=0.95\textwidth]{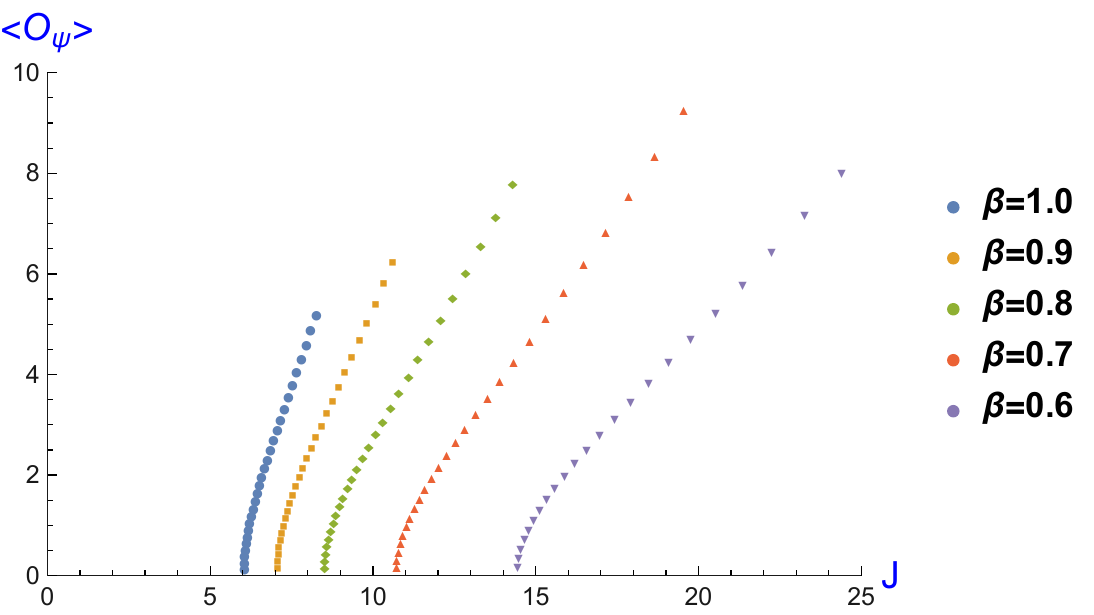}
 \caption{}\label{fig:OvsJ1}
 \end{subfigure}%
\begin{subfigure}{0.5\textwidth}
 \centering
 \includegraphics[width= 0.95 \textwidth]{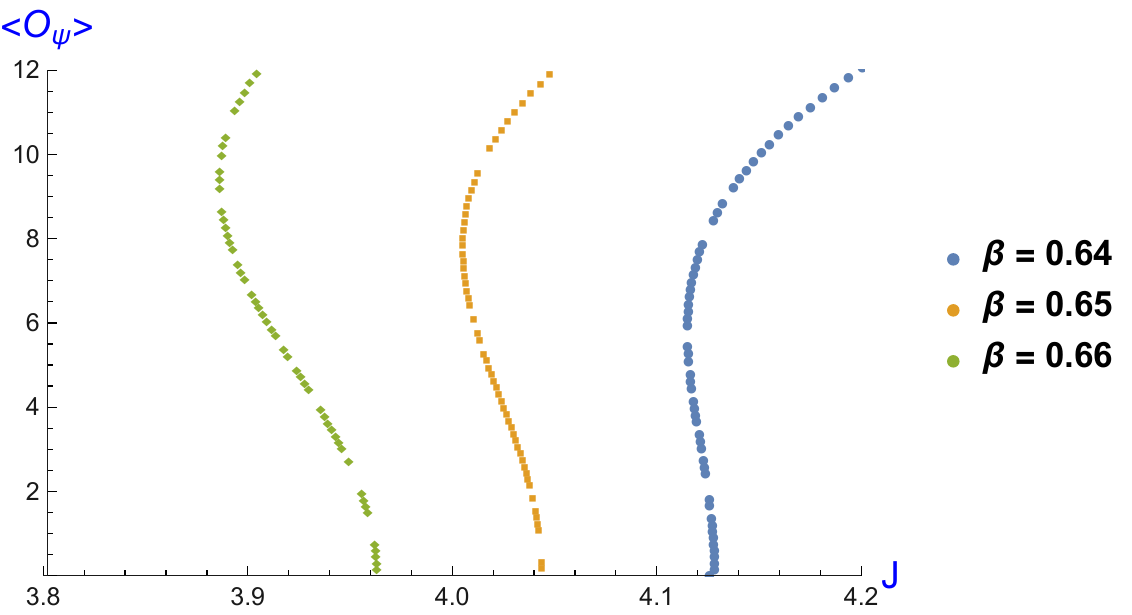}
 \caption{}\label{fig:OvsJ2}
\end{subfigure}
\caption{Plots between $\langle O_\psi \rangle$ and $J$ as we vary the strength of mutual coupling $\beta$ and $\lambda$. 
In (a), we have chosen $\lambda=0$, $\alpha_{1}=1=\alpha_2$
and the value of $\beta$ has been varied from $0.6$ to $1.0$. In (b) we have taken $\beta =0$, $\alpha_{1}=0.1 = \alpha_2$ and $\lambda = 0.64, 0.65, 0.66.$ } \label{fig:OvsJ}
\end{figure}

As we see from \eqref{expleq}, the equations of motions are a set of coupled second order ordinary 
differential equations. Therefore, we need two boundary conditions for each of the fields. The regularity of the fields 
near the horizon serve as one of the two boundary conditions. So we are to impose one more boundary condition, for each of the fields. 
This other boundary condition is imposed asymptotically, near the boundary of $AdS$. The specific values of the $J_\phi$, $J$ and $\mu$ are taken to be 
set of other boundary conditions. 

\subsubsection*{Numerical procedure}\label{ssec:numpro}

The boundary value problem, described above, is numerically solved by shooting method. In this method, we start from near the even horizon, where $\phi_0$, $\psi_0$ and $a_0$ are the 
free parameters. Then the boundary quantities $J_\phi$, $J$ and $\mu$ are all determined in terms of the near horizon free parameters. This relation is then inverted 
to obtain the desired solution. 

We perform a shooting to set the source of the charged operator $J_\psi$ to zero. This is essential to ensure that we have a phase, where the charged scalar operator has 
condensed, so as to have a superconducting phase. In practise, we first fix the value of $\psi_0$ at the horizon, and then vary the other two horizon parameters $\phi_0$ and $a_0$ to ensure 
$J_\psi=0$. This procedure ensures that we indeed have a solution, where the charged scalar field $\psi$ is non-zero. Subsequently, we vary $\phi_0$ to obtain the entire space of solutions. 
Notice that the two free parameters at the horizon (one goes away in ensuring $J_\psi=0$), maps to the two boundary free parameters $J$ and $\mu$. In this way, numerically we manipulate 
the boundary parameters $J$ and $\mu$, by varying the near horizon parameters.

A typical solution, obtained in this way, is shown in fig.\ref{fig:samsol}. This solution has been generated setting $\lambda = 0$. The $u \rightarrow 0$, region denotes the 
asymptotic AdS in our coordinates. The dotted straight lines, in the log-log plot for the scalar fields, indicates that an appropriate shooting has been performed to ensure $\phi$ scales
as $u$, while $\psi$ scales as $u^2$. This immediately implies that $J_\psi$ has been set to zero, in this solution.

\begin{figure}[H]
\centering
\begin{subfigure}{.5\textwidth}
 \centering
 \includegraphics[width=0.95\textwidth]{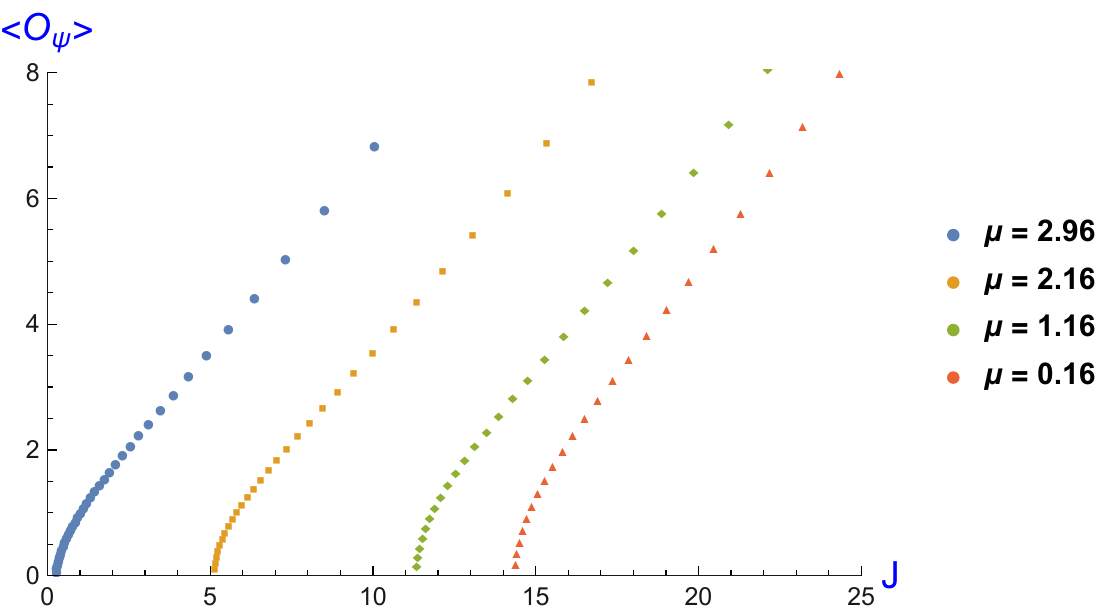}
 \caption{}\label{fig:OvsJMU1}
 \end{subfigure}%
\begin{subfigure}{.5\textwidth}
 \centering
 \includegraphics[width=0.95\textwidth]{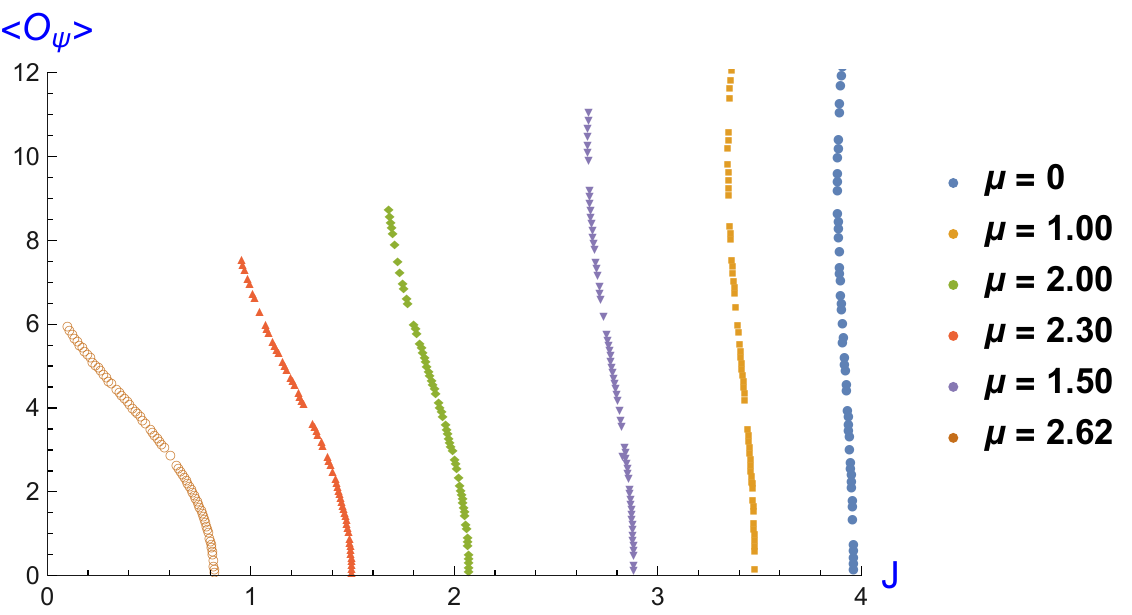}
 \caption{}\label{fig:OvsJMU2}
\end{subfigure}
\caption{ Plots for $\langle O_\psi \rangle$ and $J$ as we vary the chemical potential at fixed values of mutual interaction. In (a) we have taken $\alpha_1=1=\alpha_2$, $\lambda=0$, $\beta = 0.6$, and
$\mu$ has been varied from  0.16 to 2.96. In (b), we have taken  $\alpha_1 =0.1 = \alpha_2$, $\beta =0$, $\lambda=0.66$ and $\mu$ has been varied from 0 to 2.62. }\label{fig:OvsJMU}
\end{figure}

\subsubsection*{Case-I: $\lambda =0, \beta \neq 0$}

Let us consider the case $\lambda =0$, and hence, the mutual interaction between the two scalar fields is entirely controlled by $\beta$. Note that, although in this case, with $\lambda =0$, 
the Lagrangian enjoys a discrete symmetry of $\phi \rightarrow - \phi$, but the source of the scalar field $\phi$, definitely breaks this symmetry. 
As discussed towards the beginning of \S \ref{ssec:setup},  $\beta$ must lie within the intervals
\begin{equation}\label{betabnd}
\frac{ \alpha_2 }{8} \leq \beta \leq \sqrt{\alpha_1 \alpha_2}, ~~\text{with} ~\alpha_1 \geq 0, ~\alpha_2 \geq 0. 
\end{equation} 

At first, we consider the interesting case, when the chemical potential $\mu = 0$. Here we investigate, how $\langle \mathcal O _\psi \rangle$, the expectation value of operator dual to the charged scalar $\psi$, varies as a function of the source $J$ of the scalar field $\phi$. This has been plotted in fig.\ref{fig:OvsJ1}. 
For this plot, we have chosen $\alpha_1 =1= \alpha_2$, and the value of $\beta$ has been varied from $0.6$ to $1$. 

We notice that, for a fixed value of $\beta$, there exists a critical value of the source $J_c$, after which the condensation of the charged scalar operator $O_\psi$ takes place.
There are no superconducting phase below this value of $J_c$ and the value of $\langle \mathcal O _\psi \rangle$ increases monotonically as $J$ is increased. 
This indicates that the phase transition is second order. 
We also observe that this critical value  $J_c$ decreases, as we increase the strength of mutual interaction $\beta$. 
This suggests that the mutual interaction plays a crucial role in process of this condensation. Also, since the chemical potential has been set to zero, the external forcing of 
the uncharged scalar, via the mutual interaction is the main reason for the formation of the superconducting phase. 

At this stage, the reader may wonder, why do we call 
the phase at zero chemical potential superconducting. This is definitely a symmetry broken phase, since $\langle \mathcal O _\psi \rangle \neq 0$. The other important reason is that, 
even as we take the chemical potential to zero, this phase responds like a superconductor, to external perturbation. 
For instance, conductivity, which is the response of the system to an Electric field, behaves identically like a superconductor, as we discuss in more detail in \S \ref{ssec:linfluc}.

Next we study, how $\langle \mathcal O _\psi \rangle $ vs $J$ plot varies, as we vary the chemical potential, at fixed $\beta$. This has been plotted in fig.\ref{fig:OvsJMU1}.
We notice, that the value of $J_c$, the critical value of source of $\phi$, decreases as we increase the chemical potential. In fact, there exists a finite value of the chemical potential 
where $J_c \rightarrow 0$. This value of the chemical potential, is precisely that where the superconductor phase starts to exist in the set up of \cite{Hartnoll:2008kx}. 
In order to perform this computation with non-zero chemical potential, for the purpose of simplicity, we have first considered a charged AdSRN black hole background with the following metric and gauge field 
\begin{equation} \label{chbkg}
\begin{split}
& ds^2=\frac{1}{u^2}\left(-f(u)dt^2+\frac{1}{f(u)}du^2+dx^2+dy^2 \right),~~\text{where}~~f(u)=(1- (Q^2 +1)u^3 + Q^2 u^4 ), \\
& A = 2 Q (1-u) dt, ~~\text{with}~ \mu = 2 Q. 
\end{split}
\end{equation}
and have considered the solutions of scalar fields over this background. The qualitative nature of our results is expected to be identical, if we had worked in the  probe approximation 
described in \S\ref{ssec:setup} 
\footnote{Unlike \S\ref{ssec:setup}, this is a small charge limit, where amplitude of the scalar field 
is small, so that the scalar fields have negligible back reaction on the metric, as well as the gauge field.}.

\begin{figure}[h!]
\begin{center}
  \includegraphics[width=0.5\linewidth]{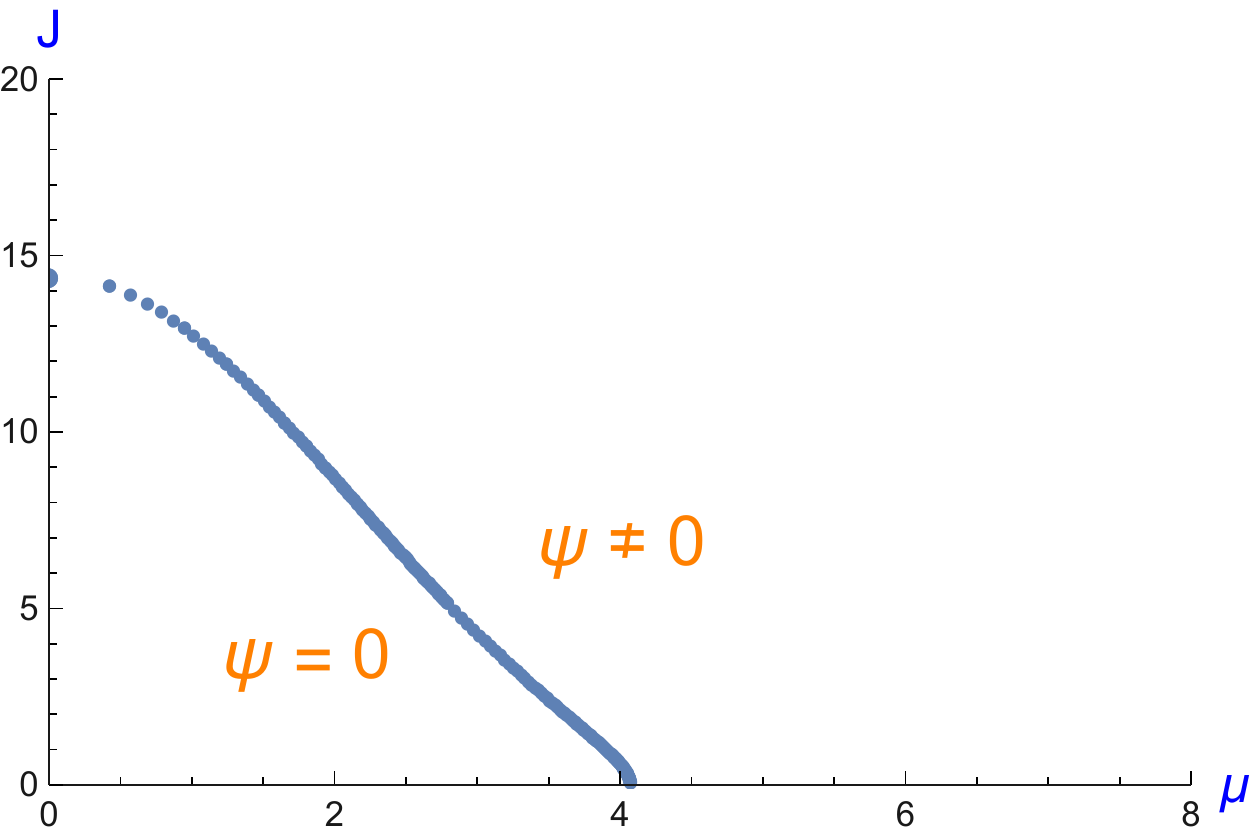}
  \caption{Phase Plot : Source of the field $\psi$ vs. chemical potential, at the point where the operator expectation value of field $\phi$ is just turned on 
  ($\langle \mathcal O _{(\phi)}\rangle \leq 0.002$). The value of the lagrangian parameters are respectively $m_1^2 = -2 = m_2^2, ~\alpha_1 =1= \alpha_2, ~\beta = 0.6$, $\lambda =0$.
  }
  \label{fig:phpl}
\end{center}
\end{figure}

Thus to summarise, we have the following scenario. If we did not have any source $J=0$, this reduces to the case 
studied in \cite{Hartnoll:2008kx}. In this case, there exists a critical value of the chemical potential $\mu_c$ beyond which 
the condensation takes place. Now as we turn on $J$ and start increasing it, this critical value of chemical potential $\mu_c$ starts decreasing. In fact, for a particular 
finite value of the source $J$, we arrive at a situation where $\mu_c$ tends to zero. The existence curve for the superconducting phase, thus has the form as shown in fig.\ref{fig:phpl}
\footnote{This plot has been obtained for the large charge limit explained in \S \ref{ssec:setup}, considering a completely dynamical gauge field. This is unlike the approximation \eqref{chbkg}, which was 
used to generate the plots in fig.\ref{fig:OvsJMU}.  }.

The phase plot in fig.\ref{fig:phpl} has a very curious feature. As discussed in \S\ref{sec:intro}, using the underlying conformal invariance of our system, this implies that the 
phase transition temperature $T_c$ at a fixed chemical potential increases as we increase $J$. This is suggestive of a possible mechanism to control the $T_c$ of superconductors. 
It would be extremely interesting to explore, if the relevant and important features of this mechanism, can be practically realised in real life high $T_c$ superconductors. 

\subsubsection*{Case-II: $\beta =0, \lambda \neq 0$}

Now, let us consider the case $\beta =0$, when the mutual interaction between the two scalar fields is entirely controlled by $\lambda$. As discussed towards the beginning of \S \ref{ssec:setup}, 
the main difference here is that due the presence of the non-zero quartic self-interactions of the scalar fields, there is no upper bound on $\lambda$. Therefore, $\lambda$ can be large compared 
to the self-couplings. 

In fig.\ref{fig:OvsJ2}, we have shown the $\langle \mathcal O _\psi \rangle $ vs $J$ plot, as we vary $\lambda$, at zero chemical potential. 
We notice that as $\lambda$ increases, the nature of the plot deviates significantly, from  
that corresponding to the lower values of $\lambda$. In particular, at the critical value of $J_c$, where the superconducting phase starts to exist, $\langle \mathcal O _\psi \rangle $ does not tend to 
zero. In fact, the value of source for which $\langle \mathcal O _\psi \rangle \rightarrow 0$, is higher than the critical value of the source $J_c$. We also note that $\langle \mathcal O _\psi \rangle $ 
as a function of $J$ is double valued for a specific range of $J$ just below $J_c$; after this range it is single valued and  increases monotonically with $J$.  This is a hallmark of a first order phase transition. 
We can confirm this by comparing the free energies of the relevant phases in \S \ref{ssec:ordphtr}. 

In fig.\ref{fig:OvsJMU2}, we show how $\langle \mathcal O _\psi \rangle $ vs $J$ plot changes, as we vary the chemical potential, keeping $\lambda$ fixed. We find that the first order nature of the 
transition becomes more manifest as we increase $\mu$. In fact, it is curious to note that, at a critical value of the chemical potential,
 there exists a solution even at zero source $J=0$, with non-zero $\langle \mathcal O _\psi \rangle$. 


No surprises are expected, if we turn on both $\beta$ and $\lambda$ simultaneously; all the interesting features has been captured in the above discussions. Hence, we refrain from a detailed discussion 
of the most general case.

\subsection{Analysis of the order of phase transition}\label{ssec:ordphtr}

In order to understand the order of the phase transition, we have to evaluate and compare the free energies in the competing phases. 
The free-energy is holographically evaluated by computing the corresponding Eucledian onshell action. The free energy $\Omega$ is related 
to the Eucledian onshell action through $\Omega = - {\mathcal S}_{os}$.

Here we have worked with the ansatz \eqref{eqan} for the given fields. 
Plugging this back into the action \eqref{realpac}, we get 
\begin{equation}
\begin{split}
\mathcal S = & T \mathcal V \int_{0}^1 du ~\left( \frac{1}{2} A_t'(u)^2 - \frac{f(u)}{u^2} \psi'(u)^2 -  \frac{f(u)}{u^2} \phi'(u)^2  \right. \\
& \left. + \frac{1}{u^4} \left( - m_1^2  \psi(u)^2 - \frac{\alpha_1}{2} \psi(u)^4 - m_2^2  \phi(u)^2 - \frac{\alpha_2}{2} \phi(u)^4 + \beta \phi(u)^2 \psi(u)^2 \right)  \right) ,
\end{split}
\end{equation} 
$\mathcal V$ being the volume in the field theory directions. 
Now performing an integration by parts on the second and on the third term we have 
\begin{equation} 
\begin{split}
\mathcal S = & T \mathcal V \int_{0}^1 du ~\left( \frac{1}{2} A_t'(u)^2 + \psi \frac{d}{du} \left( \frac{f(u)}{u^2} \psi'(u) \right) + \phi \frac{d}{du} \left( \frac{f(u)}{u^2} \phi'(u) \right)  \right. \\
& \left. + \frac{1}{u^4} \left( - m_1^2  \psi(u)^2 - \frac{\alpha_1}{2} \psi(u)^4 - m_2^2  \phi(u)^2 - \frac{\alpha_2}{2} \phi(u)^4 + \beta \phi(u)^2 \psi(u)^2 \right)  \right) \\
& - \left[ \frac{1}{u^2} f(u) \psi'(u) \psi(u) +  \frac{1}{u^2} f(u) \phi'(u) \phi(u) \right]_{u \rightarrow 0},
\end{split}
\end{equation} 	
The asymptotic expansion for the $\psi(u)$ and $\phi(u)$, for the assumed mass of the scalar fields, is given by 
\begin{equation}
\psi(u) = \langle \mathcal O_{\psi} \rangle u^2 , ~\phi(u) = J u + \langle \mathcal O_{\phi} \rangle u^2.
\end{equation}
Using this for the boundary term and using the equation of motion for the scalar fields, the onshell action is given by 
\begin{equation}
\begin{split}
\mathcal S_{os} = & T \mathcal V \int_{0}^1 du ~\left( \frac{1}{2} A_t'(u)^2 
+ \frac{1}{u^4} \left( \frac{\alpha_1}{2} \psi(u)^4 + \frac{\alpha_2}{2} \phi(u)^4 - \frac{\lambda}{2} \phi \psi^2 - \beta \phi(u)^2 \psi(u)^2 \right)  \right) \\
 &- \left[ \frac{J^2}{u} + 3 J \langle \mathcal O_{\phi} \rangle \right]_{u \rightarrow 0} + \mathcal S_{ct},
\end{split}
\end{equation} 
In order to cancel the divergent piece we need to add a mass counterterm $\mathcal S_{ct}$ for the scalar field $\phi$ 
\begin{equation}
\mathcal S_{ct} = \int_{bdy}  m_c^2 ~\phi(u^2) |_{u \rightarrow 0} =  T \mathcal V \left[ - \frac{J^2}{u} - 2 J \langle \mathcal O_{\phi} \rangle \right]_{u \rightarrow 0}, 
\end{equation} 
where we have to choose $m_c^2 = -1$, to ensure the cancellation of the divergence. Another fact that we have used here is that the square root of the induced metric 
on the boundary (a constant $u$ surface) goes as $1/u^3$. 
Using this counterterm the onshell action reduces to 
\begin{equation}
\begin{split}
\mathcal S_{os} = & T \mathcal V \int_{0}^1 du ~\left( \frac{1}{2} A_t'(u)^2 
+ \frac{1}{u^4} \left( \frac{\alpha_1}{2} \psi(u)^4 + \frac{\alpha_2}{2} \phi(u)^4- \frac{\lambda}{2} \phi \psi^2  - \beta \phi(u)^2 \psi(u)^2 \right)  \right) 
 + J \langle \mathcal O_{\phi} \rangle. 
\end{split}
\end{equation} 
As is apparent from this expression of the onshell action, when the values of the mutual interactions are comparable to the self interactions, we might expect 
that the order of phase transition changes. For smaller values of the self interactions, the phase transition remains second order. Since, situations similar to this 
has been discussed in the literature before \cite{Hartnoll:2008kx}, we refrain from going into the details of the second order phase transitions. Now, when $\lambda =0$,  $\beta$
cannot be comparable to the self couplings because of the bound arising from boundedness of the potential \eqref{betabnd}. So for all allowed values of 
$\beta$ we have a second order phase transition. However, with $\beta=0$, no such upper bound exists for $\lambda$, and the plots in 
fig.\ref{fig:OvsJ2} and fig.\ref{fig:OvsJMU2}, indicates that the order of the phase transition is first order. Hence, let us examine this case in more details.

Let us focus on the case when the gauge field is switched off, i.e. the case of zero chemical potential. In this case, there is a competition between two solutions, which 
can be schematically written as 
\begin{equation} 
\begin{split}
& \text{solution 1:}~ \psi = 0 , ~ \phi = \phi_0 ,\\
& \text{solution 2:} ~ \psi = \psi_1 , ~ \phi =  \phi_1 . 
\end{split}
\end{equation} 
Here we have simply expressed the fact that, in one of the solutions the charged scalar field $\psi$ is zero, while 
in the other one it is supported by the source of the uncharged field $\phi$. Both the solutions are to be considered at the 
same value of the source $J$. 

The difference in free energy between these solutions would therefore, be given by 
\begin{equation} 
\begin{split}
\frac{\Delta \Omega}{T \mathcal V} =  &\int_{0}^1 du ~ \frac{1}{u^4} \left( \frac{\alpha_1}{2} \psi_1(u)^4 + \frac{\alpha_2}{2} \phi_1(u)^4 - \frac{\lambda}{2} \phi(u) \psi(u)^2 - \beta \phi_1(u)^2 \psi_1(u)^2 - \frac{\alpha_2}{2} \phi_0^4 \right) \\
 &+ J \left( \langle \mathcal O_{\phi_1} \rangle - \langle \mathcal O_{\phi_0} \rangle  \right). 
 \end{split}
\end{equation}
where the fields $\psi_1$, $\phi_0$ and $\phi_1$ are solutions to the equations of motion \eqref{expleq}, in the probe limit. 

\begin{figure}[H]
\centering
 \includegraphics[width=.6\textwidth]{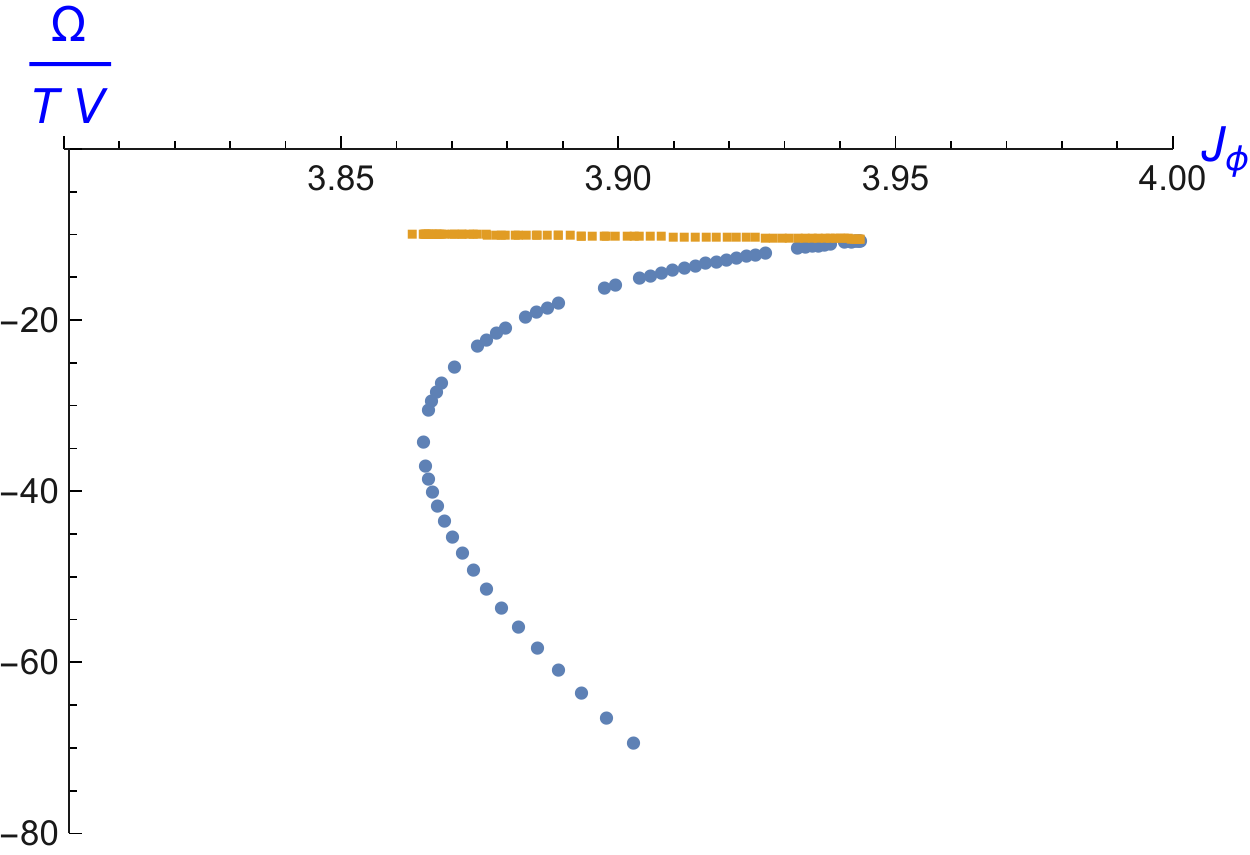}
\caption{Here we plot the difference in free energies of the two solutions, at $\alpha_1 = 0.1 = \alpha_2 $, $\beta = 0$ and $\lambda=0.66$. The orange line represents 
the normal phase, while the blue line represents the superconducting phase.}\label{fig:FE}
\end{figure}

In fig.\ref{fig:FE}, we plot the free energies of the two solutions, at zero chemical potential, for various values 
of the source $J$ of the uncharged scalar $\phi$. The orange line is the free energy of the normal phase, 
while the blue line is the free energy of the superconducting phase. At the critical value of the source $J_c$, 
where the superconducting phase starts to exist, the free energy jumps discontinuously, clearly 
demonstrating the fact that we have a first order phase transition, in this case.

\subsection{Linear fluctuations about broken phase}\label{ssec:linfluc}
In this section, we shall compute the linear response of the system, to an  electric field, through a fluctuation analysis, with the solutions 
obtained in \S\ref{ssec:numpro} as background. This linear response is given by the conductivity of the system, which can be computed using holographic techniques. 
We shall turn on the $A_x$ component of the gauge field, as a small fluctuation over the background solutions 
and determine it, by solving the Maxwell equations. From this solution, we can subsequently determine the conductivity, using the standard holographic 
prescription. 

Let us consider 
\begin{equation}
A_x = \mathcal A_x (u) \exp(- i \omega t).
\end{equation}
With this ansatz, the equation for $\mathcal A_x$ in our set up is given by 
\begin{equation}\label{Axeq}
\mathcal A_x''(u) + \frac{f'(u)}{f(u)} \mathcal A_x'(u) + \left( \frac{\omega^2}{f(u)^2} - \frac{2 \psi(u)^2}{u^2 f(u)} \right) \mathcal A_x(u) = 0, 
\end{equation}
where $f(u) = 1-u^3$, and the form of $\psi(u)$, is given by the background solution for the charged field condensate, obtained following 
the procedure outlined in \S\ref{ssec:numpro}. 

Regarding boundary conditions, we would like to set ingoing  conditions for $A_x$ at the horizon, so that we have 
\begin{equation}
\mathcal A_x = f(u)^{-\frac{i \omega}{3}} \left( c_0 + c_1 (1-u) + c_2 (1-u)^2 + \dots \right)
\end{equation}
This sets one of the boundary conditions, for the second order differential equations \eqref{Axeq}. 

\begin{figure}[H]
\centering
\begin{subfigure}{.5\textwidth}
 \centering
 \includegraphics[width=.9\textwidth]{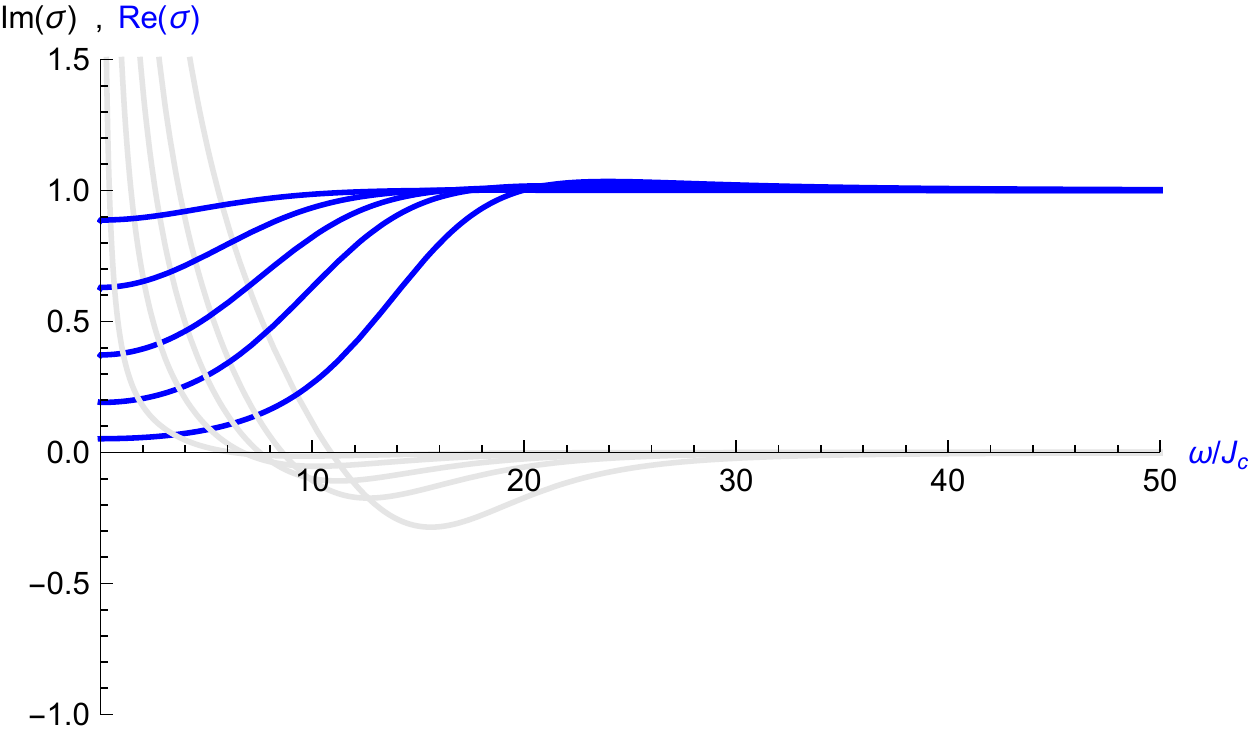}
 \caption{}\label{fig:sigvsoma}
\end{subfigure}%
\begin{subfigure}{.5\textwidth}
 \centering
 \includegraphics[width=.9\textwidth]{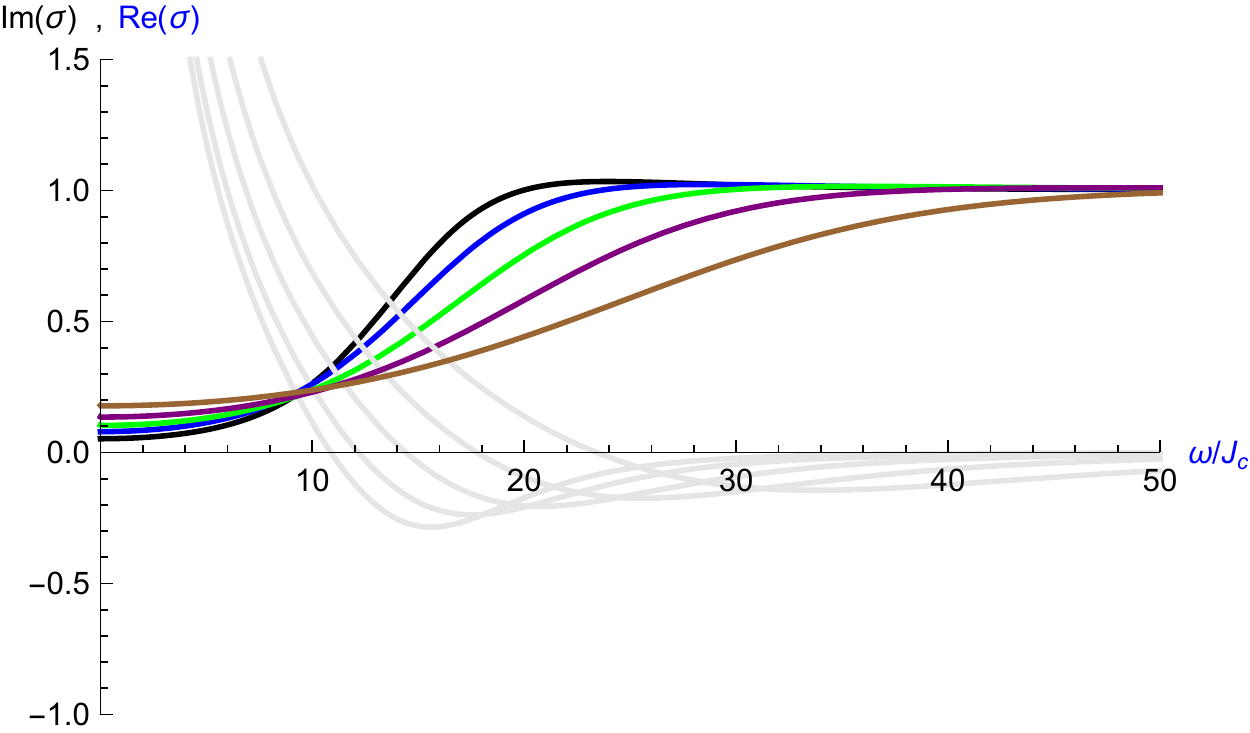}
 \caption{}\label{fig:sigvsomb}
\end{subfigure}
\caption{Plots of Real and Imaginary (light grey) part of conductivity $\sigma$ vs frequency $\omega/J_c$, at zero chemical potential, when $\lambda =0$.
Here $J_c$ is the critical value source at which the superconductor phase begins to exist. 
In (a) we have fixed the Lagrangian parameters to $\alpha_1= \alpha_2 =1$,  and $\beta = 0.6$, and we have increased $J$ starting from $J_c$. We notice that at large values of $J$,
magnitude of the gap in the conductivity plot saturates and becomes maximal. 
In (b) we have fixed the value of 
the source $J$, to some large value much above $J_c$,  where the gap is maximal, and chosen 
the Lagrangian parameters to be $\alpha_1$=1, $\alpha_2$=1, $\beta$=0.6, 0.7, 0.8, 0.9 and 1.0, from brown to black in that order. 
  }\label{fig:sigvsom}
\end{figure}

Now, asymptotically, $ A_x$ may be written as a Fefferman-Graham expansion in the radial coordinate 
\begin{equation} 
 A_x = a_x + b_x ~u + \dots,
\end{equation}
where, through the AdS-CFT dictionary, $a_x$ is related to the boundary electric field $E_x = - \dot a_x$, while $b_x$ determines 
the boundary current $J_x = b_x$. We can think of specifying $a_x$ as the second boundary condition for \eqref{Axeq}. In this way, 
once we subject the system to an electric field, at the linear level, it responds by setting up a current proportional to the electric field--Ohm's law. The 
proportionality constant gives us the conductivity 
\begin{equation}
\sigma (\omega) = \frac{J_x}{E_x} = - \frac{i b_x }{\omega a_x } 
\end{equation}

\begin{figure}[H]
\centering
\begin{subfigure}{.5\textwidth}
 \centering
 \includegraphics[width=.9\textwidth]{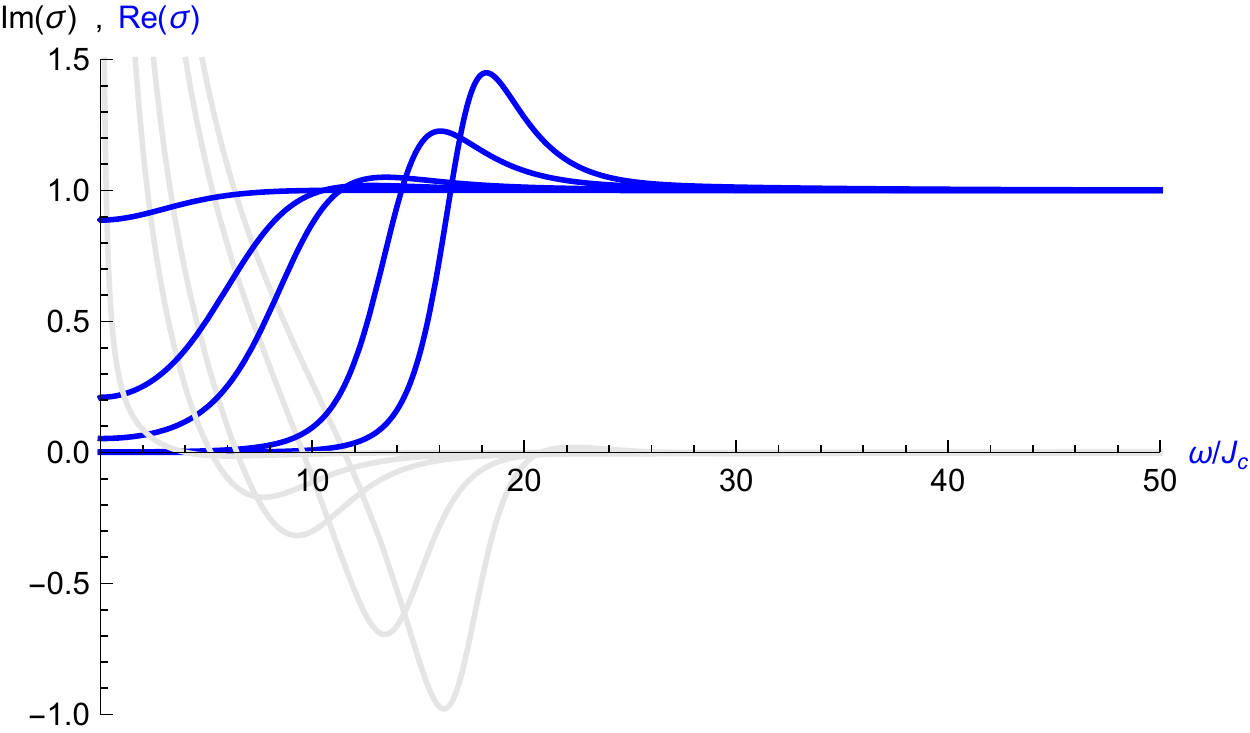}
 \caption{}\label{fig:sigvsom2a}
\end{subfigure}%
\begin{subfigure}{.5\textwidth}
 \centering
 \includegraphics[width=.9\textwidth]{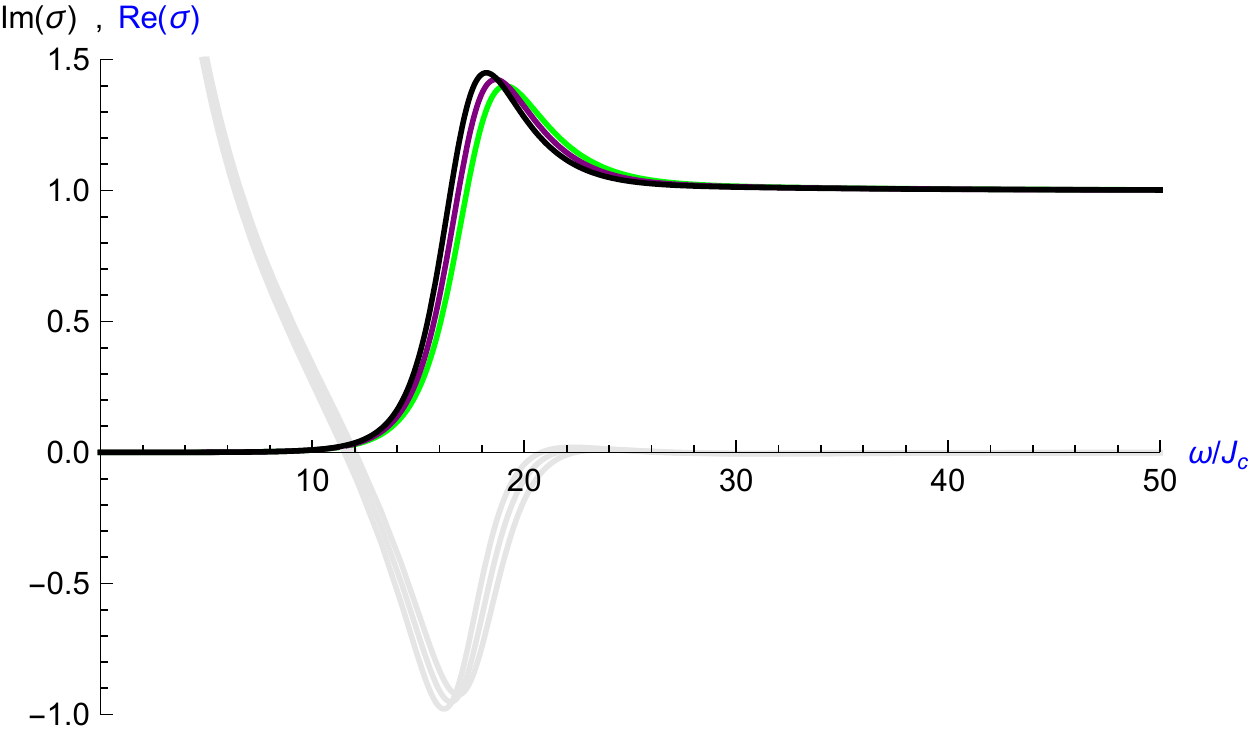}
 \caption{}\label{fig:sigvsom2b}
\end{subfigure}
\caption{ Plots of Real and Imaginary part of conductivity $\sigma$ vs frequency $\omega/J_c$, at zero chemical potential, when $\beta =0$. Here $J_c$ is the critical value source 
at which the superconductor phase begins to exist. 
In (a) we have fixed the Lagrangian parameters to $\alpha_1= \alpha_2 =0.1$,  and $\lambda = 0.66$, and we have increased $J$ starting from $J_c$. We notice that at large values of $J$,
magnitude of the gap in the conductivity plot saturates and becomes maximal. 
In (b) we have chosen the parameters  $\alpha_1$=0.1, $\alpha_2$=0.1, $\lambda$=0.64, 0.65, 0.66 and we have fixed the value of 
the source $J$, to some large value much above $J_c$,  where the gap is maximal. 
 }\label{fig:sigvsom2}
\end{figure}

The real and imaginary parts of $\sigma(\omega)$ are related, through the Kramers-Kronig relations
\begin{equation}
\begin{split}
\text{Im}[ \sigma(\omega)] 
\end{split} = -\frac{1}{\pi} \mathscr P \int_{-\infty}^{\infty} \frac{\text{Re}[\sigma(\tilde \omega)]}{\tilde \omega -  \omega} d\tilde \omega
\end{equation}

For superconductors, it is expected that the real part of conductivity has a delta function peak, near $\omega =0$, i.e. dc resistivity vanishes. 
This, in turn, implies that there is a pole in the imaginary part of conductivity, near $\omega =0$. Since, it is  significantly difficult to capture the 
delta function numerically, the presence of this delta function, is inferred by looking at the pole in the imaginary part of $\sigma(\omega)$. 

In fig.\ref{fig:sigvsom}, and fig.\ref{fig:sigvsom2} we have plotted the real and imaginary parts of $\sigma(\omega)$, for 
the superconducting phase, at zero chemical potential, with the same solutions as background, which were used to generate fig.\ref{fig:OvsJ}.
The imaginary part of conductivity clearly has a pole at $\omega =0$, confirming that we are indeed in the superconducting phase. 
Apart from the delta function, the real part of conductivity also exhibits a gap, characteristic of holographic superconductors. 

In fig.\ref{fig:sigvsoma}, and fig.\ref{fig:sigvsom2a}, we notice that the gap becomes more and more prominent as we increase the source $J$, 
and eventually saturates to a fixed value when $J$ is much above the critical value  $J_c$. 
In fig.\ref{fig:sigvsomb}, and fig.\ref{fig:sigvsom2b} the plots has been made by varying the strengths of mutual interaction ($\beta$ and $\lambda$), when the operator 
expectation value $\langle O_\psi \rangle$, for the different values of interaction strengths, are comparable and high. We find that the gap in the real of part of conductivity, 
becomes more and more shallow as we decrease the strength of mutual interaction. This again indicates that the mutual interactions play a key role, in realising 
the superconducting phase.

\section{Discussion}\label{sec:disc}

In this note, by employing holographic techniques, we have analysed the effect of interactions of the condensate in a superconducting phase, with other operators. 
We have found that the process of condensation is facilitated, if we force the coupled system with a source of the uncharged operator. This has an important effect 
of increasing the phase transition temperature $T_c$, at a given constant chemical potential. The main physical ingredient which produces this effect, is the strength  
of  interactions between the charged and uncharged operators. It may be speculated that we would continue to see this effect 
if the uncharged operator was not a scalar, but other operators with non-trivial spin. Surely, a 
concrete statement in this direction would demand further careful analysis. Admittedly, finding such an operator in a physical system would be a challenging job 
and also it may be even more challenging to establish experimental control over the source of such operators, even if they can be identified. But our work 
has a definite suggestion for a mechanism, to increase $T_c$; the viability of a physical realisation of this proposal, is left as a challenge for future work. 
It would also be worth investigating, whether such a mechanism can be used to increase $T_c$, even within the paradigm of BCS theory. 
Such a novel mechanism would be very interesting, particularly in the backdrop of recent deliberations on \cite{2018arXiv180708572T}, 
where  there has been some fascinating claims on achieving superconductivity at room temperature 
(also see \cite{2018arXiv180802929S, 2018arXiv180802005B}, for relevant discussions on the reported experimental observation). 

The novel consequence of forcing the system through the source $J$ of an interacting operator, 
and the realisation of a superconducting phase at zero chemical potential, seems to indicate an interesting fact regarding 
superconductors with strongly coupled microscopics. Our results suggest, there may be two distinct microscopic mechanisms 
at play, which leads us to the superconducting phase. One of them is controlled by the chemical potential of the charge carriers 
and the other is controlled by interactions between the charged scalar with other operators. For example, in a BCS like picture, where the cooper-pairs are formed due to electron-phonon interactions, there should be some remnant interaction between 
the cooper-pairs and the phonons, in the effective theory of the superconducting phase. 
It is extremely tempting to speculate that our uncharged scalar operator 
would be analogous to the phonons in such a picture. 
It would be very interesting to see, if this analogy can be exemplified further. 

The superconductor solution that exists as the chemical potential tends to zero is particularly interesting. In this solution, we find that the gauge field entirely vanishes and consequently, 
so does the electric field. Naively this might sound perplexing; how is it possible to have a condensate of a charged scalar field, without generating an electric field due to its charge. 
A careful examination of coupling tell us that this intuition is inaccurate. The charged scalar field can exist nontrivially, even when the gauge field is zero; the Maxwell equations 
consistently decouple from the scalar field equations in this scenario
\footnote{Note that, in the Maxwell equation in \eqref{reom}, the current sourcing the electromagnetic fields is given by $j^\mu = 2 A^\mu \psi^2$. This vanishes 
when $A^\mu =0$, even if $\psi$ has a non-zero profile. This implies that the charge density may vanish, when $\psi$ is non-zero. Also, 
since we have a gauge invariant definition of the current, $j^\mu = -i \left( \psi \left( \mathcal D_\mu \psi \right)^\star -   \psi^\star \left( \mathcal D_\mu \psi \right) \right)$,
therefore, if charge density is zero in one gauge, it remains zero for all other gauge choices. }. 
However, when there is some background charge (say, due a charged blackhole, as in our case),
 the scalar condensate modifies the gauge field, due to its non-trivial coupling with it. 

Our work also has a few other immediate direction of generalisation. Firstly it would be very interesting to work out the full back-reacted solutions and study how the source 
effects the metric components. A similar set up may also be used to generalise the standard construction of holographic superconductors. 
Such a generalisation would  be helpful to analyse the effect of this forcing, on the first order transport coefficients, appearing in the boundary energy-momentum tensor of the 
superconducting phase. It would also be very interesting to construct solutions with a spatially varying source. 
This might give interesting generalisation of the results on the striped phases reported in \cite{Ooguri:2010kt,Donos:2011bh}.
This, in turn, would give us an opportunity to study anisotropic transport properties in superconductors and superfluids, at constant chemical potential. 
Finally, it would be definitely interesting to find the right embedding of our system into supergravity, 
and determine the interaction strengths between the scalar fields in a top down approach. We postpone these questions for future investigation.


\acknowledgments 
We would like to thank Nilay Kundu, Kamal Lochan Panigrahi, Rohan Pramanick and Tirtha Sankar Ray  for several useful discussions.
We would also like to thank Nilay Kundu, for several insightful comments on the final version of this manuscript. 
 JB would like to thank ICTS, Bangalore, for hospitality during the conference 
``AdS-CFT at 20'', where this work was initiated. 
PB  wishes to thank Robert De Mello Koch and other organizers of 
``The Third Mandelstam Theoretical Physics School and Workshop'', 
during which a part of this work took shape.
JB would also like to acknowledge 
hospitality at IACS, Kolkata, during the workshop titled ``AdS/CFT correspondence and thermalisation", 
and NISER, Bhubaneswar, during the conference titled ``Mini Conference on Recent  Topics in String Theory'',
in which results from this work was presented. 

\appendix

\section{Phase transition in a toy Landau-Ginzburg model}\label{app:toy}

\begin{figure}[h!]
\begin{center}
  \includegraphics[width=0.5\linewidth]{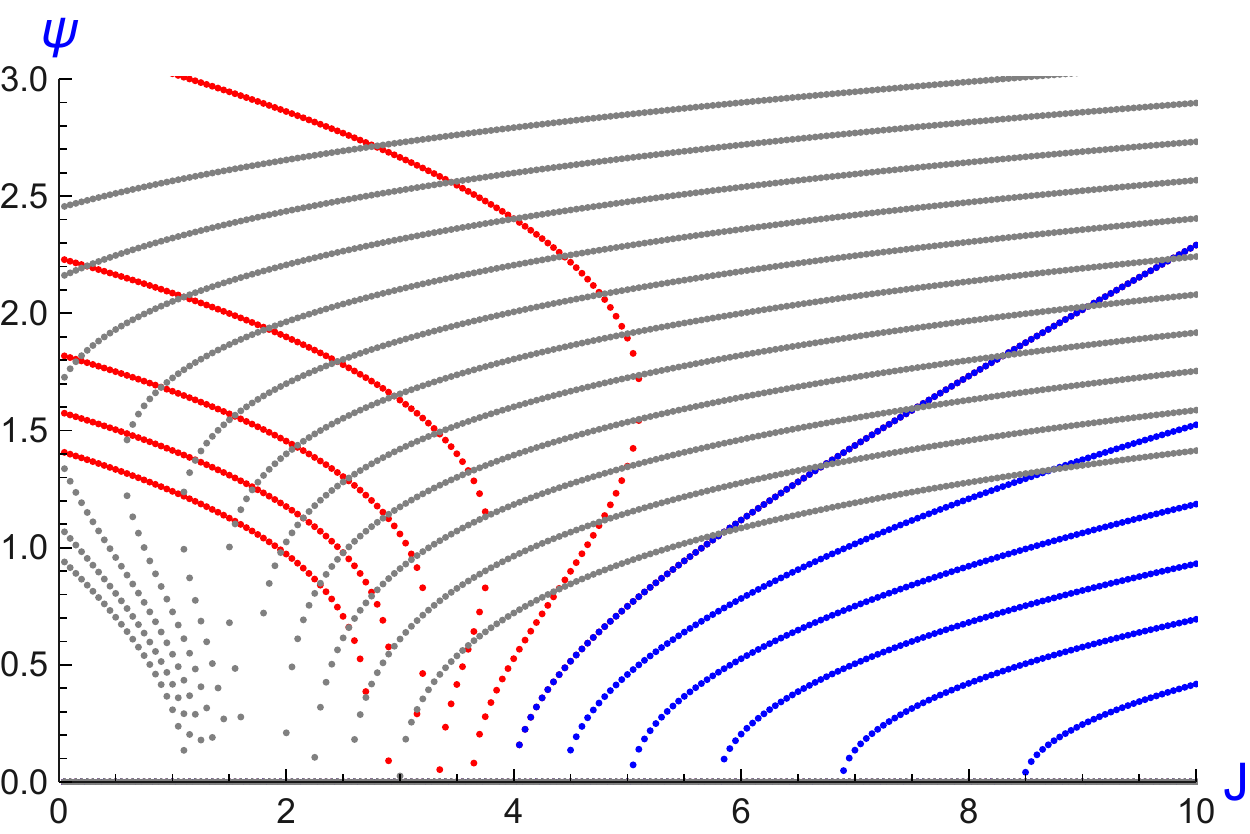}
  \caption{ 
  Plot of $\psi$ vs $J$ for the toy system \eqref{toysys}. We have chosen $m_1=m_2 = 1$. The grey lines are generated for $\beta = 0$, $\alpha_1 =0.5= \alpha_2$ and $\lambda$ is varied between 
  1.0 and 2.0. While the Red and Blue curves are generated for $\lambda =0$, $\alpha_1 =1= \alpha_2$ and $\beta$ varied between 0.5 and 1.5. The red curves are obtained when $\beta$ is greater than 1.0, which is higher than the self-interactions
  $\alpha_1$ and $\alpha_2$. 
  } \label{fig:toyOvsJ}
\end{center}
\end{figure}

Let us consider a simple Landau-Ginzburg model for two scalar fields, one of which has a non-zero source. 
\begin{equation} \label{toysys}
\mathcal L = m_1^2 \psi^2 + \frac{\alpha_1 }{2} \psi^4 + m_2^2 \phi + \frac{\alpha_2}{2} \phi^2 - \lambda \phi \psi^2 - \beta \psi^2 \phi^2  - J \phi. 
\end{equation} 
In this toy model, we can solve for $\psi$ and $\phi$ respectively, and consider only the real and positive solutions. We  study how the 
value of $\psi$ changes in these  solutions, as we change the value of source $J$  for $\phi$. We then obtain the $\psi$ vs $J$  plot, for various values of 
interaction strength $\beta$, for specific values of $m_1,m_2, \alpha_1$ and  $\alpha_2$. We present the combined plots in fig.\ref{fig:toyOvsJ}. Note the 
remarkable similarity of this plot with fig.\ref{fig:OvsJ}. Also note that as we increase the interaction strength compared to $\alpha_1$ and $\alpha_2$, 
there exists a point where the behaviour of $\psi$ vs $J$ changes drastically. 

In fig.\ref{fig:toyOvsJ}, the red and blue curves are obtained when $\lambda =0$, while the grey ones are obtained when $\beta=0$. 
The upper bound on $\beta$, arising from the requirement of the boundedness of the potential, 
 prevents the analogue of the red curves in fig.\ref{fig:toyOvsJ} from appearing in fig.\ref{fig:OvsJ}. So the phase transition is always second order for all the allowed values of $\beta$. However, 
 the analogues of the grey curves, for large value of $\lambda$ compared to $\alpha_1$ and $\alpha_2$, has been seen in fig.\ref{fig:OvsJ2}, where
 we have a first order phase transition to superconducting phase.


\bibliographystyle{JHEP}
\bibliography{Forced_superconductor}
\end{document}